\documentstyle[prl,aps,preprint,floats,epsf]{revtex}
\begin{document}

\draft
\title{Low temperature Correlation lengths of Bilayer Heisenberg Antiferromagnets and Neutron Scattering}
\author{Lan Yin and Sudip Chakravarty}
\address{Department of Physics and Astronomy, University of California, Los Angeles, CA 90095}
\date{\today}
\maketitle
\begin{abstract}
The low temperature correlation length and the static structure 
factor are computed for bilayer antiferromagnets such as YBa$_2$Cu$_3$O$_{6}$.
It is shown that energy integrated two-axis scan in neutron scattering
experiments can be meaningfully interpreted to extract correlation length in such bilayer antiferromagnets
despite intensity modulation in the momentum perpendicular to the layers. Thus,
precise measurements of the spin-spin correlation length can be performed in the
future, and the theoretical predictions can be tested. It is also shown how the same
correlation length can be measured in nuclear magnetic resonance experiments.

\end{abstract}
\newpage
Nearly thirty five years ago Birgeneau, Skalyo and Shirane\cite{Birgeneau1} invented a neutron
scattering method, known as the energy integrated two-axis scan (ETAS),
whereby the spin-spin correlation lengths of layered magnets can be measured
without explicitly measuring the dynamic structure factor for all energy and
momentum transfers and then integrating over the energy, known as the three
axis scan. Except for special circumstances, three axis scan to measure
correlation length is  difficult. 

The underlying principles of ETAS are an
analog energy integration performed by using a special geometry and the
reliance on the dominance of low energy scattering---the method is especially
effective when the critical scattering dominates. In recent years, ETAS has
served us well in unravelling the magnetic fluctuations of the parent compound
of one of the high temperature superconductors, namely the
La$_2$CuO$_4$\cite{Birgeneau2}. The experimental results are in excellent agreement
with our theoretical understanding of these two-dimensional magnets\cite{Rev}.
Unfortunately, the method does not appear to be readily extendable to a
wide class of high temperature superconductors with close magnetic
bilayers or triple layers within the unit cell; a particularly important
example is  ${ \rm YBa_2Cu_3O_{6}}$, which has a close pair of magnetic
planes within the unit cell, but it is otherwise a square lattice spin $S=1/2$ Heisenberg antiferromagnet. 
The reason for the difficulty of using ETAS is
an intensity modulation\cite{Tranquada} with the momentum transfer perpendicular to
the planes.

It is the purpose of the present paper to (a) derive the low temperature
properties of such bilayer antiferromagnets and (b) to show how ETAS can be
extended to such bilayer antiferromagnets. It is hoped that future experiments
using ETAS will be able to provide us with precise measurements of the
antiferromagnetic correlation lengths in these materials, which should be of
great help in understanding the magnetic flutuations of these bilayer
superconductors. We also briefly remark how the same correlation length can be
also measured by measuring the Cu relaxation rate in nuclear magnetic resonance
(NMR) experiments.

The Heisenberg model for a spin-$S$, nearest neighbor, square lattice, bilayer antiferromagnet is 
\begin{equation}
H=J_{\parallel}\sum_{\langle ij \rangle,p}{\bf S}_i^{(p)}\cdot{\bf S}_j^{(p)}+J_{\perp}\sum_i{\bf
S}_i^{(1)}\cdot{\bf S}_i^{(2)}.
\end{equation}
The sum in the first term is over the nearest neighbor pairs in each plane, 
where the plane index $p$ takes  two values 1 and 2. 
The second term represents the coupling between the planes. The exchange constants $J_{\parallel}$ and
$J_{\perp}$ are both positive. 

The low energy, long wavelength properties of the two dimensional  Heisenberg model is well
described by the quantum $O(3)$ nonlinear $\sigma$-model\cite{CHN}. Here, we shall consider its
generalization to coupled bilayers. The Euclidean action for this system can be written down on
general  symmetry grounds\cite{CHN}, but it can also be derived from a $(1/S)$
expansion\cite{Dotsenko}. The action is
\begin{eqnarray}
 S&=&\int^u_0 dx_0 \int d^2x \Bigg[ \sum_{p=1}^2 
	\left({1 \over 2g_0} |\partial_\mu \hat{{ \bf \Omega}}^{(p)}|^2  - h_0 \sigma^{(i)} \right) \nonumber \\
&+&
  {1 \over 2g_{\perp}^0} |\hat{{ \bf \Omega}}^{(1)}-\hat{{ \bf \Omega}}^{(2)}|^2 -{1\over2g_t^0}(\hat{{ \bf \Omega}}^{(1)}
	 \times \partial_0 \hat{{ \bf \Omega}}^{(1)}) \cdot (\hat{{ \bf \Omega}}^{(2)}
	 \times \partial_0 \hat{{ \bf \Omega}}^{(2)})\Bigg] ,
\end{eqnarray}
where $\sigma^{(p)}$ is the component of the staggered field $\hat{{ \bf \Omega}}^{(p)}$ in the direction of the
staggered magnetic field $h$. Here, all the coupling constants and dimensional variables have been scaled to
their dimensionless forms.  The label $\mu =1,2$ denotes the spatial directions, and $\mu=0$ corresponds
to the imaginary time direction. The extent in the imaginary time direction is given by $u$,
where $c_0$ is the spin-wave velocity. For mathematical purposes, it is possible to
extend the action to dimensions, $d$, other than  2 and to  the general symmetry
group $O(N)$. It is only for $d=2$ and $N=3$, however, that the model represents a
bilayer antiferromagnet. Since there are no difficulties in working directly with
$d=2$ and $N=3$, we shall use these parameters throughout the paper. For later
convenience, we define $\tilde{h}_0=h_0 g_0$, $\tilde{\gamma}_{\perp}^0={2 g_0 \over
g_\perp^0}$, and $\tilde{\delta}_t^0={g_0 \over 2 g_t^0}$.  

The relations between the $\sigma$-model parameters and the Heisenberg model parameters
are known in the large-$S$ limit\cite{Dotsenko}.
After correcting minor errors, 
the correct relations, in the limit $S\to \infty$, are
$g_0={\hbar c_0 \sqrt{1-\tilde{\delta}_t^0} \Lambda \over \rho_s^0}$,
$g_\perp^0={\hbar c_0 \sqrt{1-\tilde{\delta}_t^0} \Lambda (\Lambda a)^2 \over \rho_\perp^0}$,
$g_t^0={4 \hbar c_0 \Lambda \over \rho_\perp^0  \sqrt{1-\tilde{\delta}_t^0}}$,
and $u={\beta \hbar c_0 \sqrt{1-\tilde{\delta}_t^0} \Lambda}$,
where $\tilde{\delta}_t^0={J_\perp \over 8 J_\parallel+J_\perp}$;
the spin-stiffness constants are $\rho_s^0=J_\parallel S^2$ and $\rho_\perp^0=J_\perp S^2$, and the spin wave velocity is
$c_0={a S\over \hbar}  \sqrt{2J_\parallel(4J_\parallel+J_\perp)}$. The quantity $a$ is the lattice
constant of the Heisenberg model, and the momentum cutoff is given by $\Lambda= {\sqrt{2 \pi}\over a}$ to conserve the number of degrees of freedom.

Our calculations show explicitly \cite{IRL} that the angular momentum coupling between the layers is irrelevant for weakly
coupled bilayer systems such as  ${ \rm YBa_2Cu_3O_6}$, although it breaks the ``Lorentz invariace" and renormalizes
the spin-wave velocity. The angular momentum coupling term also reduces to a higher gradient term when the
number of layers tends to infinity. We shall omit this term from our further discussion.

This bilayer model has a massless mode and a massive mode.  We can see this by simply expanding the action to
quadratic order:
\begin{eqnarray}
S_2 &=& \int^u_0 dx_0 \int d^2x {1 \over 4g_0} \Bigg[ |\partial_\mu ({
\mbox{\boldmath$ \pi$}}^{(1)}+ {\mbox{\boldmath$ \pi$}}^{(2)})|^2
   +\tilde{h}_0 |{\mbox{\boldmath$ \pi$}}^{(1)}+{\mbox{\boldmath$ \pi$}}^{(2)}|^2 \nonumber \\
    &+& |\partial_\mu ({ \mbox{\boldmath$ \pi$}}^{(1)}-{\mbox{\boldmath$ \pi$}}^{(2)})|^2+(\tilde{h}_0
     +\tilde{\gamma}_{\perp}^0)|{ \mbox{\boldmath$ \pi$}}^{(1)}-{\mbox{\boldmath$ \pi$}}^{(2)}|^2 \Bigg],
\end{eqnarray}
where ${\mbox{\boldmath$ \pi$}}^{(i)}$ is the
transverse component of $\hat{\bf \Omega}^{(i)}$. In the absence of staggered field $\tilde{h}_0=0$, the
symmetric combination is gapless, and the antisymmetric combination has a gap $\sqrt{\tilde{\gamma}_{\perp}^0}$,
which will be called the bare dimensionless bilayer gap; the bare dimensional bilayer gap is given by
$\hbar c_0
\Lambda
\sqrt{\tilde{\gamma}_{\perp}^0}$. At finite temperatures, 
the massless mode becomes massive due to interactions, which is a topic
of this paper.

We carry out one-loop momentum-shell calculations similar to  those of Ref.~\cite{CHN} and obtain the following
renormalization group equations
\begin{eqnarray}
{dg \over dl} &=& -g+{g^2 \over 8\pi}\left[{\coth({g\over2t}\sqrt{1+\tilde{h}})\over\sqrt{1+\tilde{h}}}+
{\coth({g\over2t}\sqrt{1+\tilde{h}+\tilde{\gamma}_{\perp}})\over\sqrt{1+\tilde{h}+\tilde{\gamma}_{\perp}}}\right], \\
{dt \over dl} &=& {g t \over 8\pi}\left[{\coth({g\over2t}\sqrt{1+\tilde{h}})\over\sqrt{1+\tilde{h}}}+
{\coth({g\over2t}\sqrt{1+\tilde{h}+\tilde{\gamma}_{\perp}})\over\sqrt{1+\tilde{h}+\tilde{\gamma}_{\perp}}}\right], \\
{d\tilde{h}\over dl} &=& 2\tilde{h}, \\
\label{dhp}
{d\tilde{\gamma}_{\perp} \over dl} &=& 2\tilde{\gamma}_{\perp}-{g \tilde{\gamma}_{\perp} \over 4\pi}
{\coth({g\over2t}\sqrt{1+\tilde{h}+\tilde{\gamma}_{\perp}})\over\sqrt{1+\tilde{h}+\tilde{\gamma}_{\perp}}},
\end{eqnarray}
where $e^l$ is the length rescaling factor. The variable $t$ is the dimensionless temperature variable.
The dimensionless thickness of the slab in the imaginary time direction $u=g/t$ satisfies a simple scaling relation given by
$(g/ t)=(g_0/t_0) e^{-l}$.

\begin{figure}[htb]
\centerline{\epsfxsize=6 in \epsffile{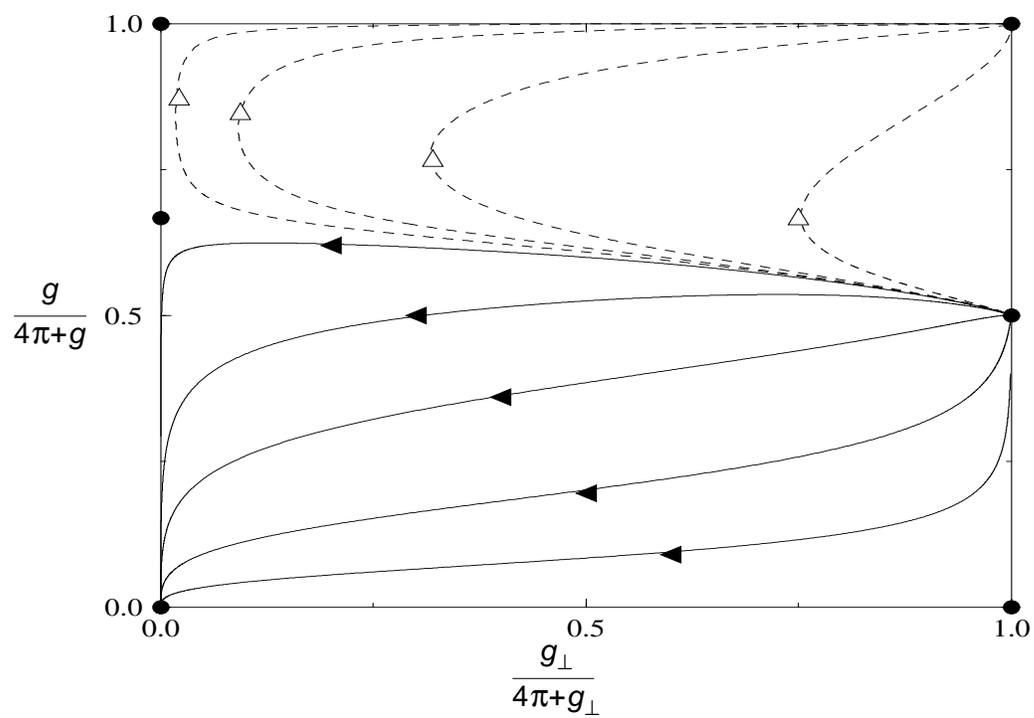}}
\caption{Zero temperature flow diagram.}
\label{ZTP}
\end{figure}

The zero temperature flows of the renormalization group equations are shown  in Fig.\ref{ZTP}. There are
two phases, separated by a separatrix between the two unstable fixed points $(g=4\pi, g_\perp=\infty)$ and
$(g=8\pi,g_\perp=0)$;  the former is the fixed point of the single layer case\cite{CHN}.  There
are two stable fixed points.  The ordered-phase fixed point is located at $(0,0)$, where both the
in- and the inter-plane couplings are infinitely strong.  The disordered-phase fixed point is located at
$(\infty,\infty)$, where the system becomes totally disordered.  Although quantum nonlinear $\sigma$-model
is a very accurate description of the low energy physics in or near the ordered phase\cite{CHN},
far into the disordered phase, such a continuum theory can not be expected to
be valid.

In the bilayer, $S=1/2$ Heisenberg model, a quantum phase transition takes place when $J_\perp \geq 2.55
J_\parallel
$\cite{BL}.  In the $\sigma$-model, the phase transtion takes place at a critical value of $g$ for any
$g_\perp$.  At present, we do not have an accurate mapping between the Heisenberg model
parameters and the $\sigma$-model parameters allowing us to relate the 
quantum disordered phases of these two models.  In the quantum disordered phase,
the large-$S$ analysis is not appropriate because it relies on the existence of
goldstone modes that are not present in this phase.

The parameters  applicable to ${ \rm YBa_2Cu_3O_{6+x}}$ lie in the ordered phase,
where $\tilde{\gamma}_{\perp}^0 \sim {J_\perp \over J_\parallel} \ll 1$, and $g$ is
below the critical value  required for the phase transition to the quantum
disordered phase at $T=0$. Therefore, the system is in the renormalized classical
regime\cite{CHN}. Also, from experiments, it is known that the ground state is an
ordered N{\'e}el state and that
$J_{\perp}\sim 0.1 J_{\parallel}$\cite{OPTM}.

To proceed further, we need an analytical solution of the renormalization group equations. We have obtained a
good appoximation to the solution  based on the following observations.
In the renormalized classical regime, the bilayer gap $\sqrt{\tilde{\gamma}_{\perp}}$ is initially much smaller than
unity, but  increases as
$\sqrt{\tilde{\gamma}_{\perp}}
\propto e^{\alpha(l) l/2}$, where $\alpha(l)<2$, but tends to  2 for large $l$.  We can
therefore consider two regions,
$\tilde{\gamma}_{\perp} \ll 1$ in region (I), and ${d\tilde{\gamma}_{\perp} \over dl }\simeq 2 \tilde{\gamma}_{\perp}$ in the 
region (II). We solve the renormalization group equations separately in regions (I) and (II), and then join
the two solutions together to get the final answer. In (I), we replace $\sqrt{1+\tilde{\gamma}_{\perp}}$
by unity and obtain
\begin{equation} \label{Temp}
{1 \over t_0}-{1 \over t_1}={1 \over 2\pi} \left[\ln \sinh({g_0 \over 2t_0}) - \ln\sinh({g_0 \over 2t_0}e^{-l_1})\right],
\end{equation}
\begin{equation} \label{Hperp}
\tilde{\gamma}_{\perp}^1={t_0 \over t_1} \tilde{\gamma}_{\perp}^0 e^{2l_1},
\end{equation}
where $(l_1,t_1,\tilde{\gamma}_{\perp}^1)$ are the variables at the endpoint of (I).  In (II), we use 
${d\tilde{\gamma}_{\perp} \over dl} = 2 \tilde{\gamma}_{\perp}$ and get
\begin{equation}
{1 \over t_1}-{1 \over t}={1 \over 4\pi} \ln \left[{\sinh({g_0 \over 2t_0} e^{-l_1} )
	\sinh({g_0 \over 2t_0} e^{-l_1} \sqrt{1+\tilde{\gamma}_{\perp}^1})
 	\over\sinh({g_0 \over 2t_0}e^{-l})
\sinh({g_0 \over 2t_0} e^{-l_1} \sqrt{e^{-2(l-l_1)}+\tilde{\gamma}_{\perp}^1})}\right].
\end{equation}

The initial condition $\tilde{\gamma}_{\perp}^0 \ll 1$ guarantees that the crossover between (I) and (II) is smooth,
namely,
$\tilde{\gamma}_{\perp}^1 \ll 1$ and ${d\tilde{\gamma}_{\perp} \over dl}|_{l_1} \simeq 2\tilde{\gamma}_{\perp}^1$.  
Since we can choose $l_1$ such that
$\tilde{\gamma}_{\perp}^0 \ll e^{-2l_1} \ll 1$, this garuantees that $\tilde{\gamma}_{\perp}^1 \ll 1$ from
Eq.~(\ref{Hperp}). The initial temperature is low enough  that both ${g_0 \over 2t_0}$ and ${g_0 \over
2t_0}e^{-l_1}$ are much larger than unity.  Following Eq.~(\ref{Temp}), we get ${1 \over t_1}\simeq{1 \over
t_0}(1-{g_0 \over 4\pi})$.  At this point, the second term in Eq.~(\ref{dhp}) becomes negligible because
${g \over 4\pi} {\coth({g\over2t}\sqrt{1+\tilde{\gamma}_{\perp}})\over\sqrt{1+\tilde{\gamma}_{\perp}}} \simeq
{g_0 t_1\over 4\pi t_0} e^{-l_1} = {g_0 \over 4\pi- g_0} e^{-l_1} \ll 2$; therefore
${d\tilde{\gamma}_{\perp} \over dl}|_{l_1} \simeq 2\tilde{\gamma}_{\perp}^1$ is also true.

The two solutions  can now be combined together to get
\begin{equation}
{1 \over t_0}-{1 \over t}={1 \over 4\pi}\ln \left[{\sinh^2({g_0 \over 2t_0}) \over \sinh({g_0 \over 2t_0}e^{-l} )
	\sinh\left({g_0 \over 2t_0}\sqrt{e^{-2l}+e^{-2l_1}\tilde{\gamma}_{\perp}^1}\right)}\right].
\end{equation}
The single layer result
\begin{equation}
{1 \over t_0}-{1 \over t}={1 \over 2\pi}\left[ \ln\sinh({g_0 \over 2t_0}) -\ln \sinh({g_0e^{-l} \over
2t_0})\right]
\end{equation}
is trivially recovered when the bilayer gap is much smaller than the inverse  correlation length, that is,
$\sqrt{\tilde{\gamma}_{\perp}^0} \ll e^{-l}$.

In the limit of large $l$, the solution has the asymptotic form of
${2 \over t}={2 \over t_{ \rm eff} } - {l \over 2\pi}$,
where
\begin{equation}
{1 \over t_{ \rm eff}} = {1 \over t_0} (1-{g_0 \over 4\pi})+{1 \over 4\pi}\ln\left[{2g_0 \over t_0}
   \sinh({g_0 \over 2t_0}\sqrt{\tilde{\gamma}_{\perp}^0(1-{g_0 \over 4\pi})})\right].
\end{equation} 
This is identical to the low temperature limit of the solution of a classical single-layer $\sigma$-model:
${1 \over t^{ \rm cl}}={1 \over t^{ \rm cl}_0}-{l \over 2\pi}$.
Thus, we can map the quantum bilayer model to the classical single layer model by identifying
${2 \over t_{ \rm eff}}={1 \over t^{ \rm cl}_0}$ and ${2 \over t}={1 \over t^{ \rm cl}}$. The remaining
arguments as to how to combine the classical two-loop result, along with the conversion of the prefactor
from the continuum to the lattice, are the same as those given in Ref.~\cite{CHN}.
We get
\begin{equation}
\xi={e \over 8} \Lambda^{-1} {t_{ \rm eff} \over 4\pi} e^{4\pi \over t_{ \rm eff}},
\label{Xi}
\end{equation}
with the exact prefactor determined in Ref.~\cite{Prefactor}.

An important result is that the argument of the exponential is twice of that of the single layer case.  This
factor of
$2$ comes from the mapping between the bilayer model and the single layer model.  To get a feeling for this
factor, it is instructive to consider a classical bilayer model given by the action
$ S={1\over t^{ \rm cl}}\int d^2x \left[ \sum_{i=1}^2 
	\left( |\partial_\mu \hat{{ \bf \Omega}}^{(i)}|^2  -\tilde{h} \sigma^{(i)} \right)
  +{\tilde{\gamma}_{\perp} \over 2} \left| \hat{{ \bf \Omega}}^{(1)}-\hat{{ \bf \Omega}}^{(2)} \right|^2 \right] $.
To map it on to the single layer model, we need to define a new field $\hat{{ \bf \Omega}}$, where the
transverse components are given by ${\mbox{\boldmath$ \pi$}}={1\over2}({\mbox{\boldmath$
\pi$}}^{(1)}+{\mbox{\boldmath$ \pi$}}^{(2)})$, and the longitudinal component is given by
$\sigma=\sqrt{1-|{\mbox{\boldmath$ \pi$}}|^2}$. This can be accomplished by integrating out the massive fields
${1\over2}({\mbox{\boldmath$ \pi$}}^{(1)}-{\mbox{\boldmath$ \pi$}}^{(2)})$. The result is the new action 
$ S'={2\over t^{ \rm cl}}\int d^2x \left(  |\partial_\mu \hat{{ \bf \Omega}}|^2  - h \sigma \right) $,
consistent with our previous result.

When temperature is low enough such that ${g_0 \over 2t_0} \sqrt{\tilde{\gamma}_{\perp}^0} >1 $, 
the correlation length can be written as 
\begin{equation}
\xi={e \over 8} {\hbar c \over 4\pi\rho_s(0)} e^{4\pi\rho_s(0) \over k_BT}\label{Xi2},
\end{equation}
where $\rho_s(0)$ is the renormalized spin stiffness constant, which  within one-loop approximation is given by
$\rho_s(0)=\rho_{s}^0 \left[1-{g_0 \over 4\pi}\left(1-{1 \over 2}\sqrt{\tilde{\gamma}_{\perp}^0(1-{g_0 \over 4\pi})}\right)\right]$.
As discussed in Ref.~\cite{CHN}, the formula can be used with $\rho_s(0)$ the exact spin stiffness constant of
the bilayer Heisenberg model. 

The numerical value of the spin stiffness constant can be obtained from spin-wave
theory, and it is given by
$
\rho_s(0) = J_\parallel S^2 Z^2_c(S) Z_\chi(S).
$
For $S=(1/2)$, $J_\perp=0.08J_\parallel$\cite{OPTM}, and up to order of $(1 /S)$,
we get $
Z_c(S)=1.153,
Z_\chi(S)=0.531,$
and $ \rho_s(0)=0.176 J_\parallel$.
Therefore the correlation length is given by
\begin{equation}
\xi \simeq 0.25 a e^{4\pi\rho_s(0) \over k_BT},
\end{equation}
where $a$ is the lattice spacing.
The exponential  temperature dependence is identical to the single layer case, but the factor
of 2 in the exponent should be noted.

The static structure factor can be obtained in the same way as in Ref.~\cite{CHN}.
For the bilayer case, it contains two pieces, one for the symmetric combination
of the unit vectors from each layer and the other for the antisymmetric combination.
The notation in the $\sigma$-model is exactly the opposite to the notation in the Heisenberg model,
since the unit vector field $\hat{{ \bf \Omega}}^{(p)}$ in the $\sigma$-model is the continuum limit of the
direction vector of the staggered spin operator $(-1)^{i+p} {\bf S}_i^{(p)}$ in the Heisenberg model. 
Our renormalization group equations show that in the long wavelength limit both pieces satisfy the following
equation as in the single layer case\cite{CHN}
\begin{equation}
S(k,t_0)=e^{2l} \left({t_0 \over t}\right)^2 S(e^lk,t),
\end{equation}
where $t\equiv t(l)$ is the running coupling constant and $k$ is defined with respect to the antiferromagnetic
wavevector
$(\pi/a,\pi/a)$. 

For the symmetric piece, we evaluate the  right hand side at
$l=l^*$ such that we are far into the disordered regime, either because the correlation length is of order
unity or because the rescaled wave vector $ke^l$ is sufficiently large. These requirements are satisfied by
choosing a $l^*$ such  that
$\xi^{-2}(t_0)e^{2l^*}+k^2 e^{2l^*}=1$. When this condition is satisfied, we can  approximate
$S_{ \rm s}[e^{l^*}k,t(l^*)]$ by the Ornstein-Zernicke form 
\begin{equation}
S_{ \rm s}[e^{l^*}k,t(l^*)] \simeq { t(l^*) \over \xi^{-2}(l^*)+k^2(l^*)}.
\end{equation}
Then, 
\begin{equation}
S_{ \rm s}(k,t_0)={t_0^2 \xi^2 \over 2 \pi} f(x),
\end{equation}
where $x=k \xi$ and $f(x)={1+{1 \over 2}\ln(1+x^2) \over 1+x^2}$.
  
The antisymmetric piece can be obtained in a similar manner.
When $l \simeq l^*$, the bilayer gap $\sqrt{\tilde{\gamma}_{\perp}}$ is increased by a factor $e^{l^*}$.
At this point, it is safe to assume the Ornstein-Zernicke form 
\begin{equation}
S_{ \rm a}(e^{l^*}k,t(l^*)) \simeq {2t(l^*) \over e^{2l^*}(\xi^{-2}_\perp+k^2)},
\end{equation} 
where $\xi_\perp$ is the length scale associated with the bilayer gap.
Approximately, we have $\xi_\perp \simeq e^{-l^*} \sqrt{\tilde{\gamma}_{\perp}} / \Lambda
\simeq\sqrt{ (1-{g_0 \over 4 \pi})\tilde{\gamma}_{\perp}^0}/\Lambda$.
In terms of the initial coupling constants, we get
\begin{equation}
S_{ \rm a}(k,t_0) \simeq {(2-{g_0 \over 2\pi}) t_0 \over k^2+\xi_\perp^{-2}}.
\end{equation}
The symmetric piece is dominant in the long-wavelength limit because
\begin{equation}
{S_{ \rm s}(0,t_0) \over S_{ \rm a}(0,t_0)} \sim t_0 {\xi^2 \over \xi^2_\perp} \gg 1.
\end{equation}

In ETAS, the wavevector of the incoming neutron ${\bf q}_i$ is fixed, while the
outgoing neutrons in a direction perpendicular to the layers are  collected,
regardless of their energies.  The transferred wavevector is given by ${\bf
q}={\bf q}_f-{\bf q}_i$.  Its in-plane component ${\bf q}_\parallel$ is a
constant,
${\bf q}_\parallel=-{\bf q}_{i\parallel}$, while its perpendicular component is
a variable, $q_\perp=q_f-q_{i\perp}$. For ${\bf q}_\parallel$ near reciprocal
lattice vectors, the form factor is approximately a constant, and the intensity
$I({ \bf q}_{\parallel})$ is proportional to 
$\int_0^\infty dq_f S^{ 3\rm D}({
\bf q}, {\hbar^2 \over 2m}(q_f^2-q_i^2))$, where $S^{ 3\rm D}({ \bf q}, \omega)$
is the 3D dynamic structure factor.  It is related to the 2D dynamic structure
factors  by $S^{ 3\rm D}({ \bf q}, \omega)= \sin^2({ q_\perp
h \over 2}) S_{ \rm a}^{ 2\rm D}({\bf q}_{\parallel},\omega)+
\cos^2({ q_\perp h \over 2}) S_{ \rm s}^{ 2\rm D}({\bf q}_{\parallel}, \omega)$,
where $h$ is the distance between the two layers. The quantity $S_{ \rm a}^{ 2\rm D}({\bf q}_{\parallel},\omega)$ corresponds to  the antisymmetric spin combination with respect to the layers, {\em symmetric} in the $\sigma$-model sense,
and $S_{ \rm s}^{ 2\rm D}({\bf q}_{\parallel},\omega)$ to the corresponding  symmetric spin combination, {\em antisymmetric} in the $\sigma$-model sense. 
In experiments one probes the region ${\bf q}_{\parallel}\approx {\bf G}$, where $\bf G$ is the nearest antiferromagnetic reciprocal lattice vector. Defining ${\bf k}={\bf q}_{\parallel}-{\bf G}$, we can rewrite the intensity in terms of the $\sigma$-mode

l structure factors, recalling that the definitions of the symmetric and the antisymmetric combinations get switched in going from the spin picture to the $\sigma$-model picture.
The intensity contains two pieces,  
$I({\bf k})=I_{ \rm s}({\bf k})+I_{ \rm a}({\bf k})$,
where
\begin{eqnarray}
I_{ \rm s} ({\bf k}) &\sim& \int_0^\infty dq_f \sin^2{(q_f-q_{i\perp}) h \over 2}
   S_{ \rm s}^{ 2\rm D{\sigma}}({\bf k}, {\hbar^2 \over 2m}(q_f^2-q_i^2)),\\
I_{ \rm a}({\bf k}) &\sim& \int_0^\infty dq_f \cos^2{(q_f-q_{i\perp}) h \over 2}
   S_{ \rm a}^{ 2\rm D{\sigma}}({\bf k}, {\hbar^2\over2m}(q_f^2-q_i^2)).
\end{eqnarray}
The $q_\perp$-modulation is unimportant in the critical region.  The reason is that
$S_{ \rm s}^{ 2\rm D{\sigma}}({\bf k},\omega)$ is dominated by the critical fluctuations near $\omega=0$,
where both $\sin^2\left({(q_f-q_{i\perp}) h \over 2}\right)$ and ${d\omega\over dq_f}={\hbar^2 \over m} q_f$ are essentially constants.
Therefore, we can pull these factors out of the integal and obtain the intensity approximately proportional to
the static structure factor, $I_{ \rm s}({\bf k}) \sim {m \over \hbar^2 q_i} \sin^2\left({(q_i-q_{i\perp})\over 2}h\right) \int_{-E_i}^{\infty} d\omega
S_{ \rm s}^{ 2\rm D{\sigma}}({\bf k},\omega) \sim S_{ \rm s}({\bf k},t_0)$, where we have reverted to the previous notation  by dropping the superscripts. The quantity $E_i$ is the incident neutron energy.
For the antisymetric piece, we get an upper bound by neglecting the factor $\cos^2({ q_\perp h \over 2}$),
which is $\int_0^\infty dq_f S_{ \rm a}^{ 2\rm D{\sigma}}({\bf k}, {\hbar^2 \over 2m}(q_f^2-q_i^2))
\sim {m \over \hbar^2 \langle q_f\rangle } S_{ \rm a}({\bf k},t_0)$,
where $\langle q_f\rangle$ is some average of the wavevector.
Because $S_{ \rm s}(0,t_0) \gg S_{ \rm a}(0,t_0)$, the intensity is dominated by the contribution
from the symmetric piece. Therefore ETAS for a bilayer should yield
information about the symmetric piece, hence the correlation length. The contribution of the antisymmetric
piece should  result in a small broad background.

The NMR relaxation rate for the in-plane Cu site will be given by\cite{CO}
${1\over T_1}\propto T^{3\over 2}\xi$, because  the symmetric structure factor 
will dominate when the correlation length $\xi$ is large. Thus, the formula for the
correlation length, Eq. (\ref{Xi2}), can also be tested in NMR experiments, as for
the single layer La$_2$CuO$_4$\cite{Imai}.

In conclusion, we have computed the low temperature properties of bilayer 
antiferromagnets using a quantum nonlinear $\sigma$-model approach. Explicit
results were given for the static structure factor and the correlation length. The
correlation length diverges exponentially as $T\to 0$, but the striking feature is
that the argument of the exponential function is a factor of 2 larger than the
corresponding single layer case. We have also shown that ETAS and NMR can be used
to test our theoretical predictions concerning the correlation length.

We thank O. Sylju{\aa}sen for discussions. This work was supported by the National Science Foundation, Grant. No. DMR-9531575.

\begin{references}
\bibitem{Birgeneau1} R. J. Birgeneau, J. Skalyo and G. Shirane, Phys. Rev. B {\bf 3}, 1736 (1971)
\bibitem{Birgeneau2} Y. Endoh {\em et al}., Phys. Rev. B {\bf 37}, 7443 (1988);
 K. Yamada {\em et al.}, Phys. Rev. B {\bf 40}, 4557 (1989).
\bibitem{Rev} For a theoretical review, see S. Chakravarty in {\em High
temperature superconductivity}, edited by K. S. Bedell {\em et al.}
(Addison-Wesley, Redwood City, 1990).For an experimental review, 
see M. Greven {\em et al.}, Z. Phys. B {\bf 96}, 465
(1995).

\bibitem{Tranquada} S. Shamoto {\em et al. }, Phys. Rev. B {\bf 48}, 13817 (1993)

\bibitem{CHN} S. Chakravarty, B. I. Halperin, and D. R. Nelson, Phys.
Rev. B {\bf 39}, 2344 (1989).
\bibitem{Dotsenko} A. V. Dotsenko, Phys. Rev. B {\bf 52}, 9170 (1995)

\bibitem{IRL} The renormalization group equation for the angular momentum coupling is given by
     $d\tilde{\delta}_t/dl=-(\tilde{\delta}_tg/4\pi) \left[ \coth \left({g \over 2t}
     \sqrt{{1+\tilde{h}+\tilde{\gamma}_{\perp} \over 1+\tilde{\delta}_t}} \right) /
     \sqrt{(1+\tilde{h}+\tilde{\gamma}_{\perp})(1+\tilde{\delta}_t)} \right].$
     Since $\tilde{\delta}_t$ always decreases and its initial value is very small,
     $\tilde{\delta}_t^0 \propto {J_\perp \over J_\parallel} \ll1$, it is irrelevant. 

\bibitem{Prefactor} P. Hasenfratz and F. Niedermayer, Phys. Lett. B {\bf 268}, 231 (1991)

\bibitem{OPTM} D. Reznik {\em et al}., Phys. Rev. B {\bf 54}, 14741 (1996); 
S. M. Hayden {\em et al.}, Phys. Rev. B {\bf 54}, 6905 (1996)

\bibitem{BL} T. Matsuda and K. Hida, J. Phys. Soc. Jpn. {\bf 59}, 2223 (1990); K. Hida, {\em ibid} {\bf 59}, 2223 (1990); {\em ibid} {\bf 61}, 1013 (1992);
	     	     A. W. Sandvik and D. J. Scalapino, Phys. Rev. Lett. {\bf 72}, 2777 (1994);
	     A. V. Chubukov and D. K. Morr, Phys. Rev. B {\bf 52}, 3521 (1995).
\bibitem{CO}S. Chakravarty and R. Orbach, Phys. Rev. Lett. {\bf 64}, 224 (1990).
\bibitem{Imai}T. Imai {\em et al.}, Phys. Rev. Lett. {\bf 70}, 1002 (1993).
\end{references}
\end{document}